\documentclass{article}

\usepackage{fullpage}
\usepackage{amssymb}
\usepackage{amsmath}
\usepackage{array}
\usepackage{lipsum}
\usepackage{nomencl}
\usepackage{pstricks}
\usepackage{tablefootnote}
\usepackage[colorlinks=true]{hyperref}
\usepackage{units}
\usepackage[utf8]{inputenc}
\usepackage{multirow}
\usepackage{dsfont}
\usepackage{authblk}
\usepackage{natbib}
\usepackage{graphicx}

\usepackage{lineno}

\newcolumntype{M}[1]{>{\centering\arraybackslash}m{#1}}
\newcolumntype{N}{@{}m{0pt}@{}}

\renewcommand{\nomgroup}[1]{%
\ifthenelse{\equal{#1}{A}}{\item[\textbf{Variables}]}{%
\ifthenelse{\equal{#1}{S}}{\item[\textbf{Subscripts}]}{
\ifthenelse{\equal{#1}{T}}{\item[\textbf{Superscripts}]}{
\ifthenelse{\equal{#1}{Z}}{\item[\textbf{Abbreviations}]}}}}}
\makenomenclature

\title{{GeoChemFoam: Operator Splitting based time-stepping for efficient Volume-Of-Fluid simulation of capillary-dominated two-phase flow}}

\author[1]{Julien Maes}
\author[1]{Hannah P. Menke}

\affil[1]{Institute of GeoEnergy Engineering, Heriot-Watt University, Edinburgh, U.K.}

\begin{document}

\maketitle



\begin{abstract}
We present a novel time-stepping method, called Operator Splitting with Capillary Relaxation (OSCAR), for efficient Volume-Of-Fluid simulations of
capillary-dominated two-phase flow. OSCAR uses operator splitting methods to separate the viscous drag and the surface tension forces.
Different time-steps are used for the viscous drag steps, controlled by the injection velocity, and for the capillary relaxation steps, controlled
by the velocity of capillary waves. Although OSCAR induces an additional numerical error of order 0 in time resulting from the splitting, it is well
suited for simulations at low capillary number. First, the splitting error decreases with the capillary number and at low capillary number,
the relaxation steps converge before reaching their last iteration, resulting in a large speed-up (here up to 250$\times$) compared to standard time-stepping
methods. The method is implemented in GeoChemFoam, our OpenFOAM\textsuperscript{\textregistered}-based CFD solver. Convergence, accuracy and efficiency are demonstrated on three benchmark cases: (1) the steady motion of an air bubble in a
straight 2D microchannel, (2) injection
of supercritical CO$_2$ in a 3D constricted channel leading to a snap-off, and (3) water drainage in a 2D oil-wet micromodel representing a porous media.
\end{abstract}







\section{Introduction}
\label{Sect:Intro}

Understanding multiphase flow at the microscale is of utmost importance for a wide range of applications, such as bio-chemical processing
\citep{2013-Elvira}, material manufacturing \citep{2006-Tavakoli}, medical diagnostic systems \citep{2014-Sackmann} and subsurface engineering processes \citep{2018-Soulaine}. Multiphase flow
involves the coupling of viscous drag in each phase with surface tension forces located at the two-fluid interface \citep{1962-Levich}. Numerical
simulations at the microscale can help investigate the fundamental physics of these processes and inform upscaling models for macroscopic applications \citep{2013-Ferrari}.

Modelling multiphase flow at the microscale is challenging due to the discontinuity of material properties at the interface between two fluids and due to interfacial boundary
conditions resulting from surface tension forces. This is particularly true at low capillary number, for which interfacial forces dominate viscous
forces locally \citep{2018-Popinet}. Accurate representation of the interface and accurate modelling of the capillary waves induced by local capillary
disequilibrium is essential.

Direct Numerical Simulation (DNS) of multiphase flow at the microscale can be achieved using the algebraic Volume-Of-Fluid
(VOF) method implemented in the OpenFOAM\textsuperscript{\textregistered} toolbox \citep{2016-OpenFOAM}, where the interface between the two fluids is captured using an indicator
function, i.e. a phase volume fraction \citep{1981-Hirt}, which is transported by solving an algebraic advection equation.
Although other methods such as geometric VOF \citep{2006-Gerlach,2009-Popinet,2014-Owkes} or level-set \citep{1994-Sussman,2018-Gibou,
2019-Luo,2020-Hashemi} can provide a more accurate description of the sharp interface, the algebraic VOF method is attractive due to its flexibility, robustness in
terms of mass conservation, and adaptability to more complex physics.  

There are two main challenges into applying DNS of multiphase flow at low capillary number. First, inaccurate computation of the interfacial force can
generate non-physical parasitic velocities \citep{1999-Scardovelli,2018-Shams,2018-Popinet}. Second, stability constraints resulting from capillary wave
velocities lead to high computational cost. Although parasitic velocities have the potential to modify the global dynamic of a two-phase system, it is
often unclear if they actually do. This is because performing simulation at very low capillary number is often not possible due to the high
computational cost. \cite{2018-Pavuluri} observed parasitic currents during numerical simulations of spontaneous imbibition in a straight microchannel
as high as six times larger than the maximum velocity expected from an analytical solution, but the error in the velocity of the two-phase interface
was less than 10\%. However, their analysis was restricted to capillary number larger than $10^{-4}$ due to the strong capillary wave stability limit,
the excessively small time-step and the large computational cost.

When simulating two-phase flow at low capillary number, the injecting velocity is low, so the time-scale of the viscous drag is large compared to the
time-scale of the capillary waves. In addition, the interface remains close to capillary equilibrium for most of the time, so the small time-steps of
the simulation are only necessary to ensure stability. Implicit and semi-implicit time-stepping, such as used in immersed boundary methods \citep{2007-Newren}, have yet to be extended to multiphase flow with surface tension forces \citep{2018-Popinet}. Alternatively, this type of problems, when two coupled operators have very different time-scales, can be
handled efficiently using Operator Splitting (OS) methods \citep{2010-Holden}.

In this paper, a novel time-stepping method, called Operator Splitting with Capillary Relaxation (OSCAR) is presented.
OSCAR uses OS to separate the viscous drag resulting from the injection velocity
and the surface tension force in the momentum equation. The total velocity and pressure fields are reconstructed using an additive OS method
\citep{2008-Farago}, and the displacement of the interface is performed using a sequential OS method \citep{2004-Carrayrou}. The viscous drag step is
solved with a time-step controlled by the injection rate and the capillary relaxation step is solved with a time-step controlled by the capillary
waves. At low capillary number, one viscous drag time-step represents a large number of capillary time-steps, but the capillary relaxation may
reach an equilibrium state so that additional steps are unnecessary, resulting in a large speed-up. Convergence, accuracy and efficiency of the method are demonstrated in three
test cases: the steady motion of an air bubble in a straight 2D microchannel, injection of supercritical CO$_2$ in a 3D constricted channel leading
to a snap-off and water drainage in an oil-wet polydisperse discs micromodel representing a porous media. 

\section{Mathematical model}
\label{Sect:model}

\subsection{Volume-Of-Fluid}

In the VOF method, the interface between two fluids is tracked using an indicator function $\alpha$, which represents the volume fraction of one of
the fluid in each grid cell. The density and viscosity of the fluid in each cell are expressed using their single-field values
\begin{linenomath}
\begin{align}
    \rho = \alpha\rho_1 + \left(1-\alpha\right)\rho_2, \\
    \mu = \alpha\rho_1 + \left(1-\alpha\right)\rho_2,
\end{align}
\end{linenomath}
where $\rho_i$ (kg/m$^3$) and $\mu_i$ (Pa.s) are the density and viscosity of phase $i$. Similarly, the velocity and pressure in the domain are expressed in term of the single-field variables
\begin{linenomath}
\begin{align}
    \mathbf{u} = \alpha\mathbf{u}_1 + \left(1-\alpha\right)\mathbf{u}_2, \\
    p = \alpha p_1 + \left(1-\alpha\right)p_2,
\end{align}
\end{linenomath}
where $\mathbf{u}_i$ (m/s) and $p_i$ (Pa) are the velocity and pressure in phase $i$. Each phase is assumed to be Newtonian and incompressible, and fluid properties are assumed to be constant in each phase.  Gravity is neglected. In this case, the single-field Navier-Stokes equations\citep{1981-Hirt} can be written as
\begin{equation}\label{Eq:NSEVOF1}
\nabla\cdot \mathbf{u}  = 0, \\
\end{equation}
\begin{equation}\label{Eq:NSEVOF2}
\rho\left(\frac{\partial \mathbf{u}}{\partial t}+ \nabla \left(\mathbf{u} \otimes \mathbf{u}\right)\right)=-\nabla p
+ \nabla\cdot\left(\mu\left(\nabla \mathbf{u} + {}^T\nabla \mathbf{u}\right)\right)+\mathbf{f}_{st},
\end{equation}
Algebraic VOF methods model the displacement of the interface by solving a discretized form of the phase advection equation.
\begin{equation}\label{phaseEq}
\frac{\partial \alpha}{\partial t} + \nabla\cdot(\alpha \mathbf{u})=0.
\end{equation}
The surface tension force $\mathbf{f}_{st}$ can be modelled using the
Continuum Surface Force (CSF) formulation introduced by \cite{1992-Brackbill}
\begin{equation}
 \mathbf{f}_{st}=\sigma\kappa\nabla\alpha,
\end{equation}
where $\sigma$ is the interfacial tension, and $\kappa$ is the mean interface curvature, which can be computed as
\begin{equation}
 \kappa = \nabla\cdot \mathbf{n}_{I},
\end{equation}
where $\mathbf{n}_{I}$ is the interface normal vector, defined as
\begin{equation}
\mathbf{n}_{I} = \frac{\nabla\alpha}{||\nabla\alpha||}.
\end{equation}

The flow regimes associated to these equations is characterised by the capillary number, which quantifies the relative importance of viscous and capillary forces,
\begin{equation}
 Ca=\frac{\mu U}{\sigma},
\end{equation}
where $U$ is the reference velocity. 


\subsection{Stability criteria}

In the VOF method, Eq. (\ref{phaseEq}) is usually solved using explicit time-stepping, i.e following
\begin{equation}\label{Eq:alphaExp}
\frac{\alpha^{n+1}-\alpha^n}{\Delta t^n}+\nabla\cdot\left(\mathbf{u}^n\alpha^n\right)=0,
  \end{equation}
where $\alpha^n$ and $\mathbf{u}^n$ are the values of $\alpha$ and $\mathbf{u}$ at time $t^n$, respectively, and $\Delta t^n=t^{n+1}-t^{n}$ is the
time-step. In addition, high resolution differencing schemes or models with compression of the interface are used to compute the
divergence of $\mathbf{u}^n\alpha^n$ in order to reduce numerical diffusion and obtain a sharp front. The surface tension force $\mathbf{f}^{n+1}_{st}$ is then calculated
using $\alpha^{n+1}$ and the pressure-velocity system defined by Eqs. (\ref{Eq:NSEVOF1}) and (\ref{Eq:NSEVOF2}) is solved for $\mathbf{u}^{n+1}$ and $p^{n+1}$ using the Pressure Implicit
Sequential Operator (PISO) algorithm \citep{1985-Issa}.
\begin{equation}
 \begin{aligned}
&\mathbf{u}^*=\frac{H^{n+1}+\mathbf{f}^{n+1}_{st}}{A^{n+1}},\\
&\nabla\cdot\left(\frac{1}{A^{n+1}}\nabla p^{n+1}\right)=\nabla\cdot\mathbf{u}^*,\\
&\mathbf{u}^{n+1}=\mathbf{u}^*-\frac{1}{A^{n+1}}\nabla p^{n+1}.
  \end{aligned}
\end{equation}
where $A^{n+1}$ and $H^{n+1}$ are the diagonal and off-diagonal elements of the coefficient matrix of Eq. (\ref{Eq:NSEVOF2}).
In all our simulations, the PISO loop is iterated three times to approach a converged solution.
The explicit time-stepping in Eq. (\ref{Eq:alphaExp}) induced a stability condition on the magnitude of the time-step, known as the
Courant-Friedrichs-Lewy (CFL) condition \citep{1928-Courant}
\begin{equation}
 \Delta t\leq \Delta t_{CFL}=\frac{\Delta x}{u},
\end{equation}
where $\Delta x$ is the distance between two grid nodes. Alternatively, implicit time-stepping can be used to avoid this time-step restriction, but
this leads to additional numerical diffusion and smearing of the sharp interface \citep{1980-Patankar}. Even for implicit time-stepping, using a
time-step larger than $\Delta t_{CFL}$ will result in difficulties to solve the pressure-velocity system \citep{1985-Issa}. In addition, the
computation of the surface tension force induced an additional time-step restriction, the so-called Brackbill condition \citep{1992-Brackbill},
which insures the stability of the capillary waves propagation
\begin{equation}
\Delta t\leq \Delta t_B=\sqrt{\frac{\overline{\rho}\Delta x^3}{2\pi\sigma}},
\end{equation}
where $\overline{\rho}$ is the average density of the two phases. At the microscale and at low capillary number, the Brackbill time-step can be
several orders of magnitude lower than the CFL time-step. To complete a simulation at the Brackbill time-step requires a very large number of
time-steps and a large computational time, but applying a larger time-step will result in non-physical results \citep{2016-Denner}. In this case we have a clear separation of time-scales between two
coupled operators in our system, the viscous drag and the capillary force. This type of problem can be handle efficiently using OS methods.

\section{Operator-Splitting based time-stepping}

The time-stepping method presented here uses OS to separate the viscous drag and the capillary forces. The domain boundaries are assumed to be
made of an inlet, with constant injected velocity, an outlet, with free-flow condition, and solid walls, with no-flow and no-slip condition. The system
is then split in two sub-systems, a viscous drag step that includes the injection velocity, which results in viscous drag velocity and pressure
$\mathbf{u}_{vd}$ and $p_{vd}$, and intermediate phase indicator function $\alpha^*$, and capillary relaxation steps that include the surface tension forces,
which results in capillary relaxation velocity and pressure $\mathbf{u}_{cr}$ and $p_{cr}$, and a new indicator function $\alpha^{**}$. Different
time-stepping strategy can be applied to the viscous and capillary relaxation steps, with the CFL condition controlling the viscous drag steps and
the Brackbill condition controlling the capillary relaxation steps. An additive OS method \citep{2008-Farago} is used for the velocity and pressure,
i.e.
\begin{equation}
 \begin{aligned}
  &\mathbf{u}=\mathbf{u}_{vd}+\mathbf{u}_{cr},\\
  &p=p_{vd}+p_{cr},
 \end{aligned}
\end{equation}
while a sequential OS method \citep{2004-Carrayrou} is used for the phase function, i.e.
\begin{equation}
\begin{aligned}
&\alpha^*\left(t^n\right)=\alpha^n,\\
&\alpha^{**}\left(t^n\right)=\alpha^*\left(t^{n+1}\right)\\
&\alpha^{n+1}=\alpha^{**}\left(t^{n+1}\right)
\end{aligned}
\end{equation}
The advantage of this formulation is that at low capillary number, the capillary relaxation steps may converge before reaching their final step,
leading to a large reduction of CPU time.

\subsection{Viscous drag step}
The first step of our solution procedure is to solve the viscous drag step. Given the viscous drag velocity
$\mathbf{u}_{vd}^n$ and pressure $p_{vd}^n$, and the phase indicator function $\alpha^n$ at time $t^n$, the intermediate phase indicator function
$\alpha^*$ is given by 
\begin{equation}\label{Eq:VOFd}
\begin{aligned}
  &\frac{\alpha^*-\alpha^n}{\Delta t^n}+\nabla\cdot\left(\mathbf{u}^n_{vd}\alpha^n\right)=0
  \end{aligned}
  \end{equation}
with boundary conditions
\begin{equation}
\begin{aligned}
  &\nabla\alpha^* = 0 \hspace{0.3cm}  \text{on } \partial\Omega_{inlet}\cup\partial\Omega_{outlet}, \\
  &\frac{\nabla\alpha^*}{\|\nabla\alpha^*\|}\cdot \mathbf{n}_{wall}=\cos\theta \hspace{0.3cm}\text{on } \partial\Omega_{wall},
  \end{aligned}
  \end{equation}
The viscous drag velocity $\mathbf{u}^{n+1}_{vd}$ satisfies the momentum conservation equation
\begin{equation}\label{Eq:NSE1d}
\begin{aligned}
  &\frac{\rho^{n+1} \mathbf{u}^{n+1}_{vd}-\rho^{n} \mathbf{u}^{n}_{vd}}{\Delta t^n}+\nabla\cdot\left(\rho^{n+1} \left(\mathbf{u}^n_{vd}
  +\mathbf{u}^n_{cr}\right)\otimes \mathbf{u}^{n+1}_{vd}\right)=-\nabla p^{n+1}_{vd}
  +\nabla\cdot\mu^{n+1}\left(\nabla \mathbf{u}^{n+1}_{vd}+{}^T\nabla\mathbf{u}_{vd}^{n+1}\right)
  \end{aligned}
  \end{equation}
with the mass conservation equation
\begin{equation}\label{Eq:md}
 \nabla\cdot \mathbf{u}^{n+1}_{vd}=0
 \end{equation}
and boundary conditions
\begin{equation}
\begin{aligned}
  &\mathbf{u}^{n+1}_{vd}=\mathbf{u}_{inlet}\hspace{0.3cm}  \text{and } \nabla p^{n+1}_{vd}=0 \hspace{0.3cm}   \text{on } \partial\Omega_{inlet}, \\
  &\nabla\cdot\mathbf{u}^{n+1}_{vd}=0\hspace{0.3cm}   \text{and } p^{n+1}_{vd}=0 \hspace{0.3cm}  \text{on  }  \partial\Omega_{outlet}, \\
  &\mathbf{u}^{n+1}_{vd}=0 \hspace{0.3cm}  \text{and  } \nabla p^{n+1}_{vd}=0 \hspace{0.3cm}  \text{on }\partial\Omega_{wall}.
  \end{aligned}
\end{equation}
The phase indicator equation (Eq. (\ref{Eq:VOFd})) is solved with explicit time-stepping, while 
the pressure-velocity system defined by Eq. (\ref{Eq:NSE1d}) and Eq. (\ref{Eq:md}) is solved
using the PISO algorithm \citep{1985-Issa}. 
The viscous drag step is solved with a time-step $\Delta t_{vd}<\Delta t_{CFL}$ to ensure stability. 

\subsection{Capillary relaxation steps}
After the viscous drag step, capillary relaxation is performed. In order to ensure stability of the capillary waves, the global time-step
$\Delta t^n=t^{n+1}-t^n$ is split into $N$ sub-steps $\Delta t^{n,k}=\Delta t^n/N<\Delta t_B$. For each sub-step, the phase indicator function
is evolved first following
\begin{equation}\label{Eq:VOFc}
\begin{aligned}
  &\frac{\alpha^{**,k+1}-\alpha^{**,k}}{\Delta t^{n,k}}+\nabla\cdot\left(\mathbf{u}^{n,k}_{cr}\alpha^{**,k}\right)=0 
   &\alpha^{**,0}=\alpha^{*}\\
  \end{aligned}
  \end{equation}
with boundary conditions
\begin{equation}
\begin{aligned}
  &\nabla\alpha^{**,k+1} = 0 \hspace{0.3cm}  \text{on } \partial\Omega_{inlet}\cup\partial\Omega_{outlet}, \\
  &\frac{\nabla\alpha^{**,k+1}}{\|\nabla\alpha^{**,k+1}\|}\cdot \mathbf{n}_{wall}=\cos\theta \hspace{0.3cm}\text{on } \partial\Omega_{wall},
  \end{aligned}
  \end{equation}
The surface tension force $f_{st}^{n,k+1}$ is then computed using the updated value of the phase indicator function.  
The capillary relaxation velocity $\mathbf{u}^{k+1,n}_{cr}$ satisfies momentum conservation equation
\begin{equation}\label{Eq:NSEc}
\begin{aligned}
  &\frac{\left(\rho\mathbf{u}_{cr}\right)^{n,k+1}-\left(\rho\mathbf{u}_{cr}\right)^{n,k}}{\Delta t^{n,k}}+\nabla\left(\rho^{n,k+1} \left(\mathbf{u}^{n}_{vd}
  +\mathbf{u}_{cr}^{n,k}\right)\otimes \mathbf{u}_{cr}^{n,k+1}\right)=-\nabla p^{n,k+1}_{cr}
  + \nabla\cdot\mu^{n,k+1}\left(\nabla \mathbf{u}^{n,k+1}_{cr}+{}^T\nabla \mathbf{u}^{n,k+1}_{cr}\right)
  +\mathbf{f}^{n,k+1}_{st} \\
  &\mathbf{u}^{n,0}_{cr}\left(t^{n}\right)=\mathbf{u}^{n}_{cr}\left(t^{n}\right),
  \end{aligned}
  \end{equation}
with the mass conservation equation
\begin{equation}\label{Eq:mc}
 \nabla\cdot \mathbf{u}^{n,k+1}_{cr}=0
 \end{equation}
and boundary conditions
\begin{equation}
\begin{aligned}
  &\mathbf{u}^{n,k+1}_{cr}=0\hspace{0.3cm}  \text{and } \nabla p^{n,k+1}_{cr}=0 \hspace{0.3cm}   \text{on } \partial\Omega_{inlet}, \\
  &\nabla\cdot\mathbf{u}^{n,k+1}_{cr}=0\hspace{0.3cm}   \text{and } p^{n,k+1}_{cr}=0 \hspace{0.3cm}  \text{on  }  \partial\Omega_{outlet}, \\
  &\mathbf{u}^{n,k+1}_{cr}=0 \hspace{0.3cm}  \text{and  } \nabla p^{n,k+1}_{cr}=0 \hspace{0.3cm}  \text{on }\partial\Omega_{wall}.
  \end{aligned}
\end{equation}
Again, the phase indicator equation (Eq. (\ref{Eq:VOFc})) is solved with explicit time-stepping, while 
the pressure-velocity system defined by Eq. (\ref{Eq:NSEc}) and Eq. (\ref{Eq:mc}) is solved
using the PISO algorithm \citep{1985-Issa}. 
The residual of the $k^{th}$ sub-steps
is defined as the initial residual of the first PISO iteration
\begin{equation}
 E_k=\|\nabla\cdot\left(\frac{1}{A^{n,k+1}_{cr}}\nabla \right)\cdot p^{n,k}_{cr}-\nabla\cdot\mathbf{u}_{cr}^*\|
\end{equation}
where $\|.\|$ is the normalized norm
\citep{2016-OpenFOAM}. The relaxation steps are stopped at $k_f$ if $k_f=N$ or if $E_{k_f}<10^{-4}$. Then, the solution is moved to the next global
time-step using
\begin{equation}
 \begin{aligned}
 &\mathbf{u}^{n+1}_{cr}=\mathbf{u}^{n,k_f}_{cr},\\
 &p_{cr}^{n+1}=p_{cr}^{n,k_f},\\
 &\alpha^{n+1}=\alpha^{**,k_f},
  \end{aligned}
\end{equation}
and the global variables
\begin{equation}
 \begin{aligned}
 &\mathbf{u}^{n+1}=\mathbf{u}_{vd}^{n+1}+\frac{1}{N}\sum_{k=1}^{k_f}\mathbf{u^{n,k}_{cr}},\\
 &p^{n+1}=p_{vd}^{n+1}+p_{cr}^{n+1}.
 \end{aligned}
\end{equation}

\section{Implementation}

\begin{figure}[!b]
\begin{center}
\includegraphics[width=0.45\textwidth]{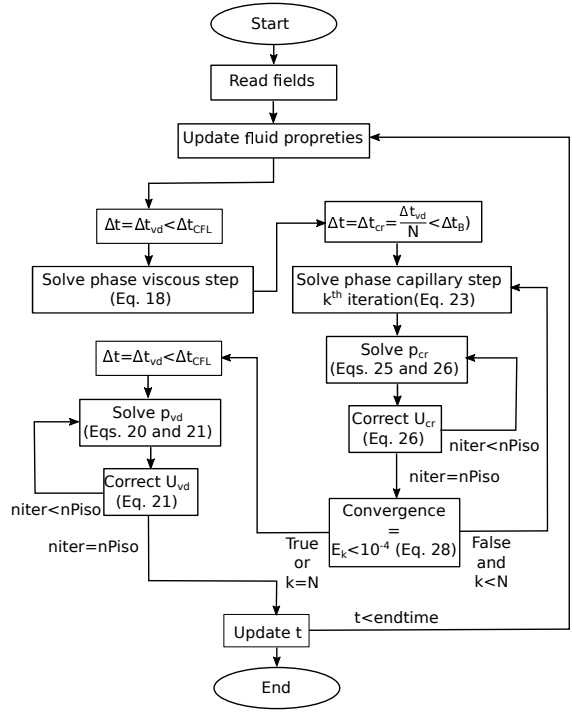}
\caption{Full solution procedure for \textit{interOSFoam}\label{fig:solutionProcedure}.}
\end{center}
\end{figure}

The numerical method has been implemented in GeoChemFoam 
(\href{https://julienmaes.com/geochemfoam}{https://julienmaes.com/geochemfoam}),
our OpenFOAM\textsuperscript{\textregistered}-based  \citep{2016-OpenFOAM} reactive transport solver. 
The standard VOF solver of OpenFOAM\textsuperscript{\textregistered}, so-called \textit{interFoam}, has been modified for this purpose into another
solver called \textit{interOSFoam}. The full solution procedure is presented in Fig. \ref{fig:solutionProcedure}.
The divergence in the momentum equation (Eq. (\ref{Eq:NSEVOF2})) is discretized using linear upwinding, while the divergence in the phase fraction equation (Eq. (\ref{phaseEq})) is
discretized using the second-order \textit{vanLeer} scheme \citep{1974-vanLeer}. To limit the smearing of the interface, an artificial compression velocity is introduced, following the method implemented in OpenFOAM\textsuperscript{\textregistered} \citep{2016-OpenFOAM}.

\section{Benchmark cases}

Convergence, accuracy and efficiency of OSCAR are evaluated on three benchmark cases. The convergence is assessed on the first test case
using the relative $L_2$ error, defined as
\begin{equation}\label{L2}
 Err=\max_{t, \beta=U, p, \alpha}\sqrt{\frac{\int{\left(\beta\left(x,t\right)-\beta_{ref}\left(x,t\right)\right)^2dx}}
 {\int{\beta^2_{ref}\left(x,t\right)dx}}},
\end{equation}
where the reference values are the ones obtained at mesh resolution of 1 $\mu$m. 
The $L_2$ splitting error induced by OSCAR is defined as
\begin{equation}\label{L2split}
 Err_{split}=\max_{t, \beta=U, p, \alpha}\sqrt{\frac{\int{\left(\beta_{\text{OSCAR}}\left(x,t\right)-\beta_{\text{PISO}}\left(x,t\right)\right)^2dx}}{\int{\beta^2_{\text{PISO}}\left(x,t\right)dx}}}.
\end{equation}

\subsection{Benchmark case 1: bubble motion in a straight 2D channel}
The first benchmark case considered is the steady motion of a non-wetting air bubble through a straight two-dimensional channel. An air bubble
($\rho=1$ kg/m$^3$ and $\mu=18$ $\mu$Pa.s) is initially at capillary equilibrium in a 2D straight channel of length 600 $\mu$m and width $100$ $\mu$m
in ethanol ($\rho=789$ kg/m$^3$ and $\mu=1.2$ mPa.s). The interfacial tension is equal to 20 mN/m and the contact angle $\theta$ is set at $0^o$ toward the wetting phase (the liquid phase). The bubble is initialised as a rectangle of
length $L=200$ $\mu$m and width $100$ $\mu$m, at a position $x=20$ $\mu$m from the left boundary, and is relaxed until capillary equilibrium. Then,
at t=0, we inject from the left boundary ethanol at constant velocity $U$ ranging from 0.167 m/s ($Ca=10^{-2}$) to 1.67~$\mu$m/s ($Ca=10^{-7}$).

First convergence of OSCAR is investigated. Simulations with different mesh resolution are performed for each capillary number.
In each case, the viscous drag steps are performed with a time-step $\Delta t_{vd}=0.01 \Delta t_{CFL}$ and 
the capillary relaxation steps are performed with a time-step $\Delta t_{cr}=\Delta t_{vd}/N$, $N$ being the smallest integer for which
$\Delta t_{cr}<\Delta t_B$.
Table \ref{Table:L2error} gives the relative $L_2$ error (Eq. \ref{L2}) at dimensionless time $t^*=Ut/L=1.0$ for all capillary numbers, and for mesh resolution
$\Delta x=\Delta y$= 2, 5 and 10 $\mu$m. The order 1 convergence behaviour is clearly visible.
\begin{table}[!ht]
\renewcommand\arraystretch{1.5}
\centering
\begin{tabular}{c|c|c|c|c|c|c}
\hline
 Resolution & $Ca=10^{-2}$ & $Ca=10^{-3}$ & $Ca=10^{-4}$ & Ca=$10^{-5}$ & $Ca=10^{-6}$ & $Ca=10^{-7}$ \\
  \hline
 2 $\mu$m & 0.019 & 0.016 & 0.015 & 0.015 & 0.015 & 0.015 \\
 5 $\mu$m & 0.061 & 0.054 & 0.054 & 0.054 & 0.054 & 0.054 \\
 10 $\mu$m & 0.083 & 0.078 & 0.079 & 0.079 & 0.079 & 0.079 \\
 \end{tabular}
 \caption{$L_2$ errors (Eq. \ref{L2}) of OSCAR for simulation of air bubble motion in a straight 2D microchannel at different capillary numbers and different mesh
 resolutions. The order 1 convergence behaviour is clearly visible. \label{Table:L2error}}
\end{table}

\begin{figure}[!t]
\begin{center}
\includegraphics[width=0.99\textwidth]{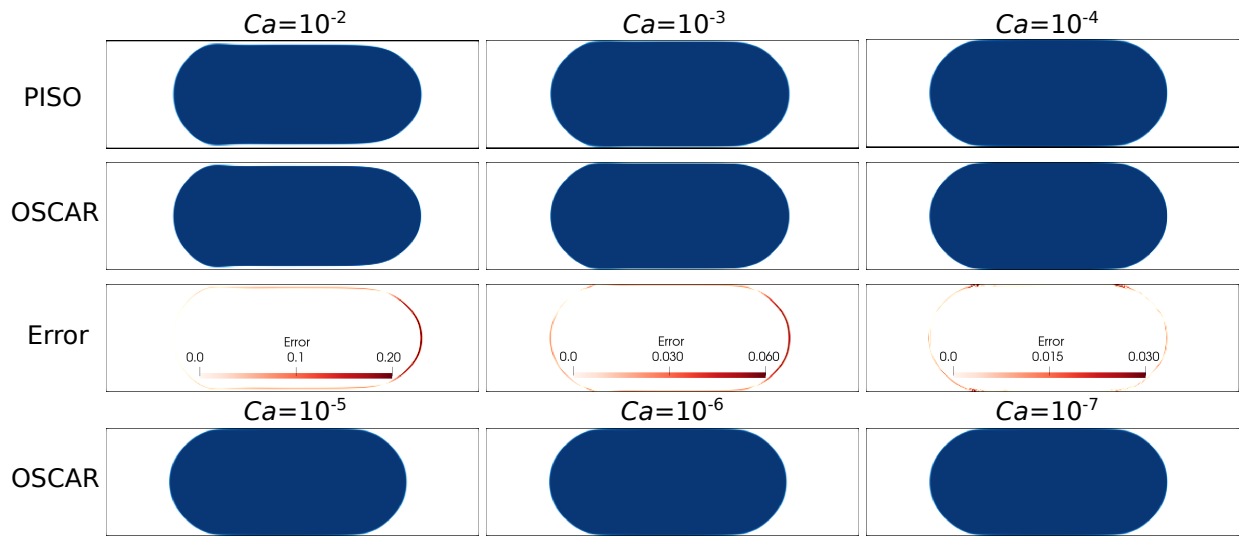}
\caption{Bubble shape at $t^*=Ut/L=1.0$ during simulation of air bubble motion in a straight two-dimensional channel at capillary numbers from $10^{-2}$ to $10^{-7}$. The simulations for $Ca\geq10^{-4}$ are performed with PISO and OSCAR, and the local splitting error in phase volume fraction is shown. The simulations for $Ca<10^{-4}$ cannot be performed with PISO due to high computational cost and only the OSCAR results are shown.
\label{fig:bubble}}
\end{center}
\end{figure}

\begin{figure}[!t]
\begin{center}
\includegraphics[width=0.85\textwidth]{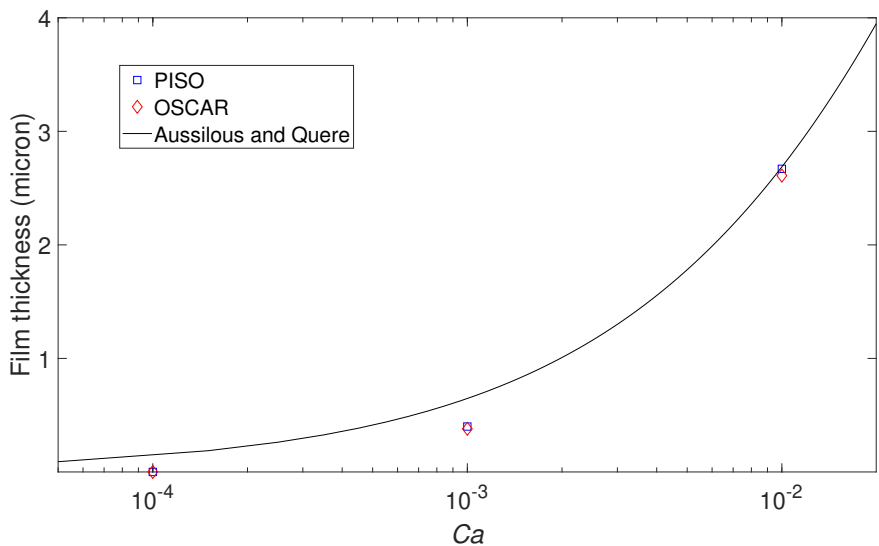}
\caption{Comparison of thin film thickness during steady motion of an air bubble in a 2D channel at various capillary numbers obtained by numerical simulations with PISO and OSCAR and with experimentally validated correlation by
\cite{2000-Aussillous}. \label{fig:filmThickness}}
\end{center}
\end{figure}

In order to compare accuracy and CPU time of OSCAR with PISO, simulations on the reference grid at all capillary numbers and various CFL are performed up to a dimensionless time $t^*=~Ut/L=~1.0$. 
Fig. \ref{fig:bubble} shows the bubble shape at dimensionless time $t^*=Ut/L=1.0$ obtained with PISO and OSCAR with CFL~$=~10^{-4}$, for $Ca=~10^{-2}, 10^{-3}$ and $10^{-4}$. 
The film is fully resolved for $Ca=10^{-2}$, partially resolved for $Ca=10^{-3}$ and unresolved for $Ca\leq 10^{-4}$. The calculated film thickness is presented in Fig. \ref{fig:filmThickness} and compared with the experimentally
validated correlation by \cite{2000-Aussillous}. At $Ca=10^{-2}$, the film thickness predicted by the correlation is 2.69 $\mu$m and we obtain 2.67 $\mu$m with PISO and 2.61 $\mu$m with OSCAR, which corresponds to errors of 0.7\% and 3\%. At $Ca=10^{-3}$, the correlation gives a thickness of 0.65 $\mu$m, and we obtain 0.39 $\mu$m with PISO (error 40\%) and 0.38 $\mu$m (error 42\%). We conclude that OSCAR is capable of modelling the thin film thickness with similar accuracy than PISO.  

The local splitting error in the phase volume fraction is also shown on Fig. \ref{fig:bubble}. For $Ca=10^{-2}$ and $Ca=10^{-3}$, the splitting error is mostly localized on the right part of the bubble. This is due to the slightly smaller film thickness, resulting in a slightly lower bubble velocity.
Table \ref{Table:BubbleSplit} shows the evolution of the L$_2$ splitting error (Eq. \ref{L2split}) as a function of the capillary and the CFL numbers. For $Ca<10^{-2}$,
not all simulations could be perform with PISO due to the Brackbill stability limit. This occurs when $\Delta t_{CFL}>\Delta t_B$. The simulations
that could not be performed are marked with ``$-$". The order 0 behaviour
with respect to time (i.e. CFL) is clearly visible. However, the error remains low for all simulations. The maximum error occurs for $Ca=10^{-2}$
and CFL $=10^{-2}$ and is equal to $1.4\times 10^{-2}$. We also observe that the error reduces with respect to $Ca$ with an order of about 0.5.

\begin{table}[!t]
\renewcommand\arraystretch{1.5}
\centering
\begin{tabular}{c|c|c|c}
\hline
 $Ca$ & CFL=10$^{-2}$ & CFL=10$^{-3}$ & CFL=10$^{-4}$ \\
  \hline
 $10^{-2}$ &  $1.4\times 10^{-2}$     & $1.3\times 10^{-2}$     & $1.2\times 10^{-2}$    \\
 $10^{-3}$ &  $-$      & $4.2\times 10^{-3}$     & $4.1\times 10^{-3}$     \\
 $10^{-4}$ &  $-$      & $-$      & $8.9\times 10^{-4}$    \\
 \end{tabular}
 \caption{Splitting error as a function of the capillary and the CFL numbers \label{Table:BubbleSplit}}
\end{table}

\begin{table}[!b]
\renewcommand\arraystretch{1.5}
\centering
\begin{tabular}{c||c|c|c|c||c|c|c|c}
\hline
    & \multicolumn{4}{c||}{PISO} & \multicolumn{4}{c}{OSCAR} \\
    \hline
        & \multicolumn{4}{c||}{CFL} & \multicolumn{4}{c}{CFL} \\
 $Ca$ & 10$^{-2}$ & 10$^{-3}$ & 10$^{-4}$ & 10$^{-5}$ & 10$^{-2}$ & 10$^{-3}$ & 10$^{-4}$ & 10$^{-5}$\\
  \hline
 $10^{-2}$ &  \color{blue}0.48 & 3.00 & 19.0 & 163$^*$ & 0.89               & 5.96               & 36.1                  & 290$^*$ \\
 $10^{-3}$ &  $-$  & \color{blue}3.25 & 19.9 & 168$^*$ & 4.01 & 6.12               & 38.1                  & 299$^*$ \\
 $10^{-4}$ &  $-$  & $-$  & \color{blue}20.0 & 181$^*$ & 26.6 & 21.4 & 38.2                  & 291$^*$  \\
 $10^{-5}$ &  $-$  & $-$  & $-$  & 214$^*$ & 24.3 & \color{blue} 21.5 &  38.3    & 293$^*$  \\
 $10^{-6}$ &  $-$  & $-$  & $-$  & $-$     &  24.4 & \color{blue} 21.5 & 38.3    & 292$^*$  \\
 $10^{-7}$ &  $-$  & $-$  & $-$  & $-$     &  24.4 & \color{blue} 21.6 & 38.3 & 293$^*$\\
 \end{tabular}
 \caption{Comparison of CPU time between PISO and OSCAR for the simulation of the steady motion of a air bubble in a 2D channel at various 
 CFL and capillary numbers.\label{Table:CPU}}
\end{table}

The computational cost of these simulations is presented
in Table \ref{Table:CPU}. Simulations with CFL $=10^{-5}$ could not be performed due to high computational time, but in order to have an estimation of
the simulation time for $Ca=10^{-5}$ using PISO, simulations with CFL $=10^{-5}$ were performed until a dimensionless time $t^*=Ut/L=0.1$ and the total
computational cost was extrapolated. For each capillary number, the minimum computational time is presented in blue.

For $Ca=10^{-2}$, the CPU time using OSCAR is roughly twice as large as the CPU time using PISO. For all these simulations,
$\Delta t_{CFL}<\Delta t_B$, and OSCAR performs one viscous drag step and one capillary relaxation step for each global time-step. For $Ca<10^{-2}$, the simulations
that could not be performed with PISO due to the Brackbill stability limit are marked with ``$-$". These simulations can be performed using OSCAR thanks to the different time-steps for viscous drag and
capillary relaxation. This is of no interest for $Ca\geq 10^{-4}$, since the minimum CPU time for PISO, obtained with CFL=$Ca$, is lower than the
minimum CPU time for OSCAR. However, at $Ca=10^{-5}$, the simulation with PISO could only be performed at CFL $=10^{-5}$, and the (extrapolated) CPU
time is 214 hours. Although the simulation with OSCAR at CFL $=10^{-5}$ would be even longer, for CFL $\leq 10^{-5}$ the capillary relaxation steps
converge rapidly. OSCAR is therefore able to provide a converged results in 24.3 hours using CFL $=10^{-2}$ or 21.5 using CFL $=10^{-3}$, roughly 10
time faster than what can be achieved using PISO. In addition, reducing the capillary number further by a factor 10 or 100 has little impact on the
CPU time of OSCAR and simulations for $Ca\leq10^{-5}$ can be performed. The bubble shape at dimensionless time $t^*=Ut/L=1.0$ for $Ca=10^{-6}$, $Ca=10^{-6}$ and $Ca=10^{-7}$ is also shown in Fig. \ref{fig:bubble} but only with OSCAR, as these simulations are impossible to perform with PISO due to high computational time. 

\begin{figure}[!b]
\begin{center}
\includegraphics[width=0.85\textwidth]{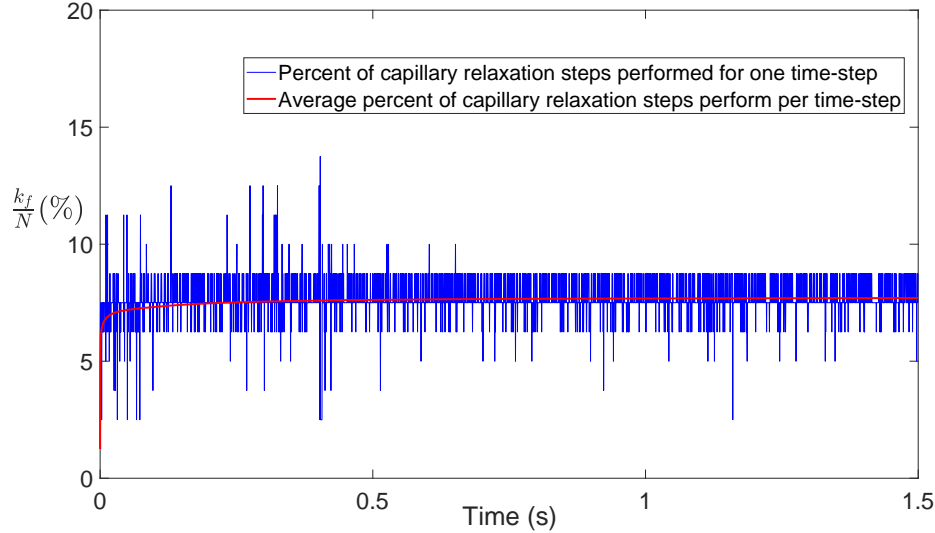}
\caption{Percent of relaxation steps performed for each time-step and average percent of relaxation steps performed as a function of time during steady motion of an air bubble in a 2D channel at $Ca=10^{-5}$.\label{fig:relaxStepsBubble}}
\end{center}
\end{figure}

In order to understand the origin of the speed-up obtained, the percent of relaxation steps performed for each time-step and average percent of relaxation steps performed
for $Ca=10^{-5}$ and CFL=$10^{-3}$ are plotted in Fig. \ref{fig:relaxStepsBubble}. In this case, one viscous drag step correspond to $N=80$ capillary relaxation step. When the relaxation steps converged for $k_f<N$, a speed-up is obtained. For most of the time-steps, the capillary relaxation converges in between 4 and 8 steps, which corresponds to 5 and 10\% of $N$. The average number of relaxation steps is 6.2, which corresponds to 7.75\% of $N$, and provides a large speed-up   

We conclude that OSCAR is able to simulate accurately the motion of an air bubble in a 2D microchannel at all capillary numbers considered here, from
$Ca=10^{-2}$ to $10^{-7}$, and is particularly efficient for $Ca\leq 10^{-4}$ with low splitting error.

\subsection{Benchmark case 2: snap-off in a 3D constricted channel}
The second benchmark case consider a 3D constricted channel with a square cross section. The geometry is made of two cubes of side $L=300$ $\mu$m separated by a square-shaped tube of side 100 $\mu$m and length 300 $\mu$m in the middle. The inlet is located at the left boundary and the outlet at the right boundary. The other boundaries are solid. The domain is meshed with a Cartesian grid of resolution 5 $\mu$m that contains 456,000 cells.

\begin{figure}[!t]
\begin{center}
\includegraphics[width=0.7\textwidth]{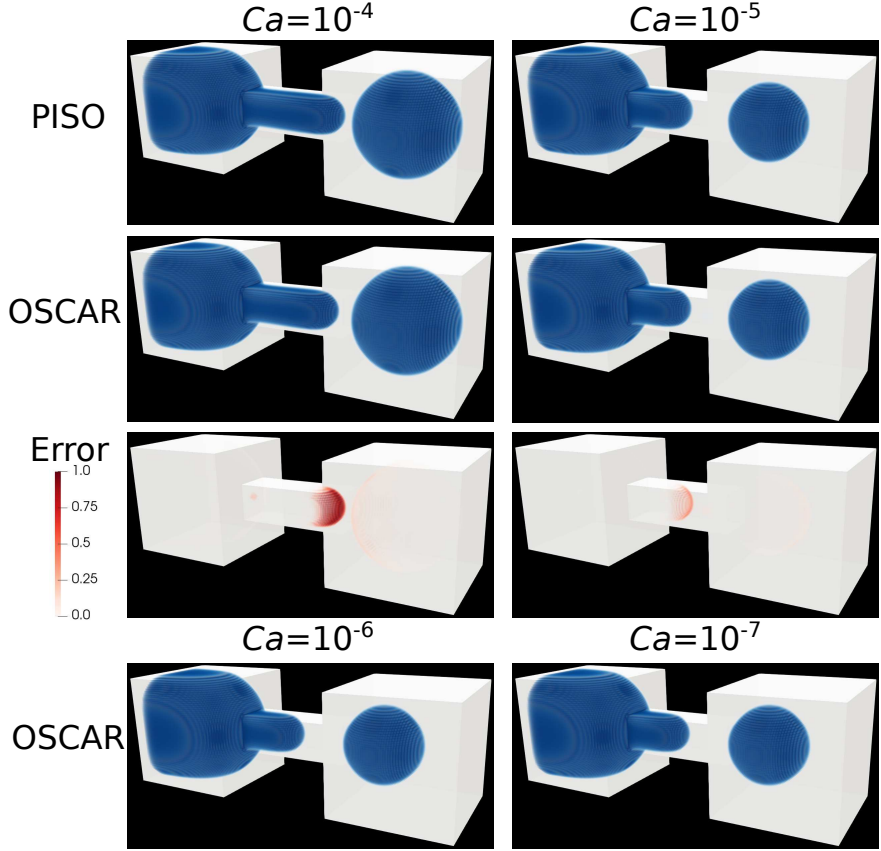}
\caption{Phase distribution after snap-off during injection of supercritical CO$_2$ (blue) in pore-constriction. The simulations for $Ca\geq10^{-5}$ are performed with PISO and OSCAR, and the local splitting error in phase volume fraction is shown. The simulations for $Ca<10^{-5}$ cannot be performed with PISO due to high computational cost and only the OSCAR results are shown.\label{fig:snapOff}}
\end{center}
\end{figure}

The domain is initially filled with water at 40$^oC$ and 9 MPa ($\rho=996$ kg/m$^3$ and $\mu=0.65$ mPa.s). At t=0, we inject supercritical CO$_2$ ($\rho=485.5$ kg/m$^3$, $\mu=0.035$ mPa.s and $\sigma=26$ mN/m) from the left boundary at constant velocities corresponding to capillary numbers $Ca=10^{-4}, 10^{-5}, 10^{-6}$ and $10^{-7}$. The contact angle $\theta$ is set at $0^o$ toward the wetting phase (the water). Since the aspect ratio $A=3>2$, the injection will lead to a snap-off in the constriction \citep{2017-Roman} and a CO$_2$ bubble will form (Fig. \ref{fig:snapOff}). Each simulation is performed using PISO with a constant
time-step $\Delta t_{PISO}=\min\left(0.1\Delta t_{CFL},\Delta t_B\right)$ and using OSCAR with a constant time-step $\Delta t_{OSCAR}=0.1\Delta t_{CFL}$. Each simulation is run until snap-off occurs unless mentioned otherwise.

Fig. \ref{fig:snapOff} shows the phase distribution after snap-off for $Ca=10^{-4}$ and $Ca=10^{-5}$ using PISO and OSCAR. The local splitting error in the phase volume fraction is also shown in Fig. \ref{fig:snapOff}. Most of the error is located at the interface after the snap-off occurred. The splitting leads to a slight delays in the snap-off and as a result the bubble is slightly larger.
\begin{table}[!b]
\renewcommand\arraystretch{1.5}
\centering
\begin{tabular}{c||c|c||c|c||c|c||c}
\hline
 Ca & \multicolumn{2}{c||}{CPU time (hour) } & \multicolumn{2}{c||}{Snap-off time (s)} & \multicolumn{2}{c||}{Bubble volume (m$^3$)}
 & Splitting error \\
    & PISO & OSCAR & PISO & OSCAR & PISO & OSCAR &  \\
  \hline
 $10^{-4}$ & \color{blue} 6.8 & 13.3 & 0.028  & 0.028  & 8.54$\times10^{-12}$  & 8.66$\times10^{-12}$ &  0.0804\\
 $10^{-5}$ & 30.9 & \color{blue} 13.7 & 0.14 & 0.14 & 3.70$\times10^{-12}$& 3.75$\times10^{-12}$ &  0.0166 \\
 $10^{-6}$ & 300$^*$ & \color{blue} 24.6  & $-$ & 1.4 & $-$  & 3.72$\times10^{-12}$  & $-$ \\
 $10^{-7}$ & 3000$^*$ & \color{blue} 24.6  & $-$ & 14 & $-$  & 3.72$\times10^{-12}$  & $-$ \\
 \end{tabular}
 \caption{Comparison of CPU time, snap-off time and bubble volume obtained with PISO and OSCAR and evolution of the splitting error
 (Eq. (\ref{L2split})) for the simulation of snap-off in a 3D constricted channel at various capillary numbers.\label{Table:constriction}}
\end{table}
Table \ref{Table:constriction} shows the CPU time, snap-off time and bubble volume obtained with PISO and OSCAR and the evolution of the splitting error with respect to the capillary number. The simulations at $Ca=10^{-6}$ and $Ca=10^{-7}$ with PISO could not be performed until snap-off due to high CPU time and the ones reported in the table were obtained by extrapolation. These simulations could be performed using OSCAR and the phase distribution after snap-off is shown on Fig. \ref{fig:snapOff}.
The snap-off time and bubble volume obtained with PISO and OSCAR are similar and the difference decreases with $Ca$. In addition, the splitting error decreases with $Ca$, with an order of approximately 0.7. OSCAR gives a speed-up compared to PISO for capillary number $Ca\leq10^{-5}$. OSCAR is 2.2 time faster than PISO for $Ca=10^{-5}$, 12 time faster for $Ca=10^{-6}$ and 120 time faster for $Ca=10^{-7}$.

\begin{figure}[!t]
\begin{center}
\includegraphics[width=0.85\textwidth]{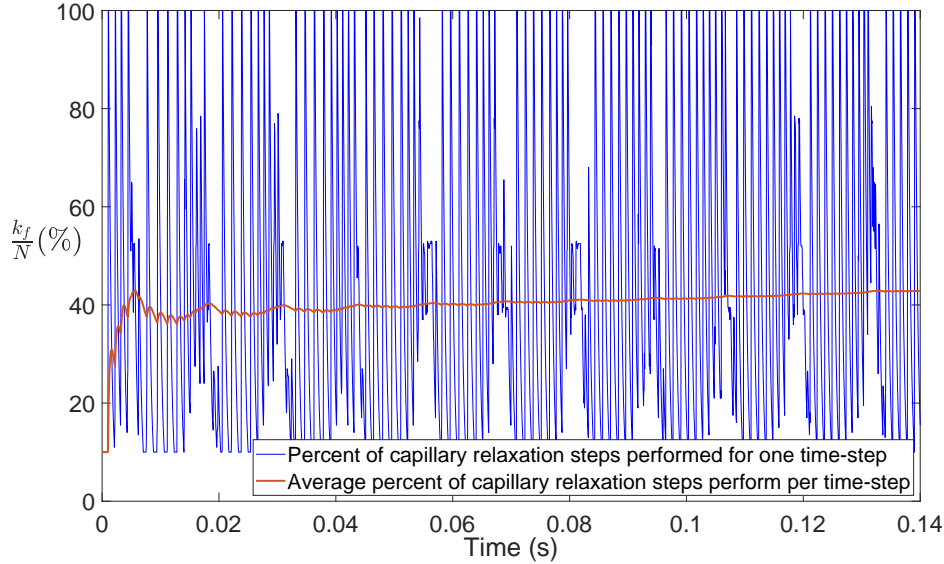}
\caption{Percent of relaxation steps performed for each time-step and average percent of relaxation steps performed as a function of time during snap-off in a 3D constricted channel at $Ca=10^{-5}$.\label{fig:relaxStepsConstriction}}
\end{center}
\end{figure}

The percent of relaxation steps performed for each time-step and average percent of relaxation steps performed
for $Ca=10^{-5}$ are plotted in Fig. \ref{fig:relaxStepsConstriction}. When the relaxation steps converged before $k_f=N$, a speed-up is obtained. However, when $k_f$ reaches $N$ without convergence, then 100\% of the relaxation steps have been performed and no speed-up is obtained. The absence of convergence can have two causes. First, the viscous drag might be too strong for the relaxation steps to converge within a time corresponding to $k_f<N$. In this case, the interface is not at capillary equilibrium for the time-step considered. Second, the absence of capillary equilibrium in the numerical simulation can be caused by numerical errors, e.g. parasitic currents. In average 42\% of the capillary relaxation steps are performed. However, the oscillatory nature of the number of capillary relaxation steps in Fig. \ref{fig:relaxSteps} suggests that parasitic currents play a large role in the convergence of our numerical solution and reducing them could further improve the CPU time of OSCAR.

We conclude that OSCAR is capable of simulating snap-off in a 3D constricted channel with similar accuracy as PISO, and that for $Ca<10^{-5}$ the simulation can be performed in approximately 24 hours, while these simulations cannot be performed with PISO due to high CPU time. 

\subsection{Water drainage in an oil-wet micromodel}

\begin{figure}[!t]
\begin{center}
\includegraphics[width=0.45\textwidth]{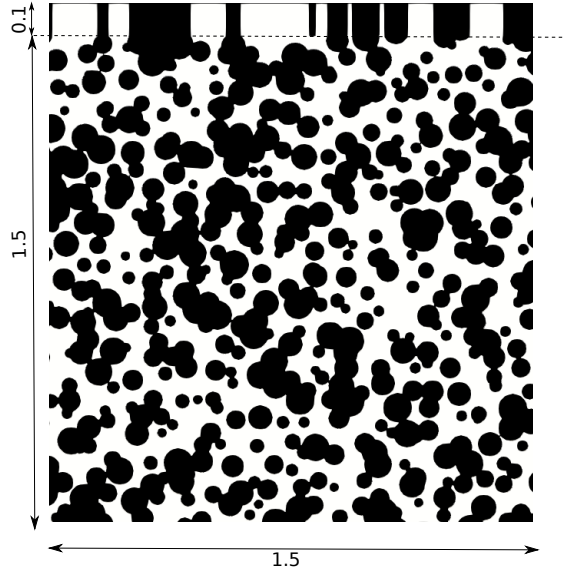}
\caption{Micromodel geometry. Dimensions are in cm. \label{fig:HM12_6_12}}
\end{center}
\end{figure}

The final benchmark case considers a 2D micromodel made of polydisperse discs representing a porous media. The model is constructed from a homogeneous domain with discs radius 270 $\mu$m by adding a random deviation of magnitude 270 $\mu$m in disc radius and center position, following the method presented in \citep{2021-Patsoukis} and the code is available open source (\href{https://github.com/hannahmenke/drawmicromodels}{https://github.com/hannahmenke/DrawMicromodels}). The geometry is presented in Fig. \ref{fig:HM12_6_12}. At the inlet situated at the top, the pores have been prolonged by tubes of 0.1 cm, so that the injected fluid can be relaxed to capillary equilibrium before the injection starts. 

To mesh the domain, a 2D uniform Cartesian grid comprising 600$\times$500 grid blocks of size 30 $\mu$m is generated, and then all cells containing solid are
removed and replaced by hexahedral and tetrahedral cells that match the solid boundaries using the snappyHexMesh OpenFOAM\textsuperscript{\textregistered} utility \citep{2016-OpenFOAM}.
The final grid contains 141,600~cells.

The domain is initially filled with dodecane ($\rho=750$ kg/m$^3$ and $\mu=1.8$ mPa.s). At t=0, water is injected ($\rho=1000$ kg/m$^3$, $\mu=1$ mPa.s and $\sigma=50$ mN/m.)
from the inlet at constant velocities corresponding to capillary numbers from $Ca=10^{-2}$ to $Ca=10^{-7}$. We chose a relatively high contact angle of 45$^o$ (weak drainage). Each simulation are performed using PISO with a constant
time-step $\Delta t_{PISO}=\min\left(0.1\Delta t_{CFL},\Delta t_B\right)$ and using OSCAR with a constant time-step
$\Delta t_{OSCAR}=0.1\Delta t_{CFL}$. Unless mentioned otherwise, each simulation is run until 5 pore volumes have been injected, i.e when
\begin{equation}
 \frac{UA}{V}t=5
\end{equation}
where $U$ is the injecting velocity, $A$ the area of the inlet and $V$ the total pore-volume of the domain.

\begin{figure}[!t]
\begin{center}
\includegraphics[width=0.85\textwidth]{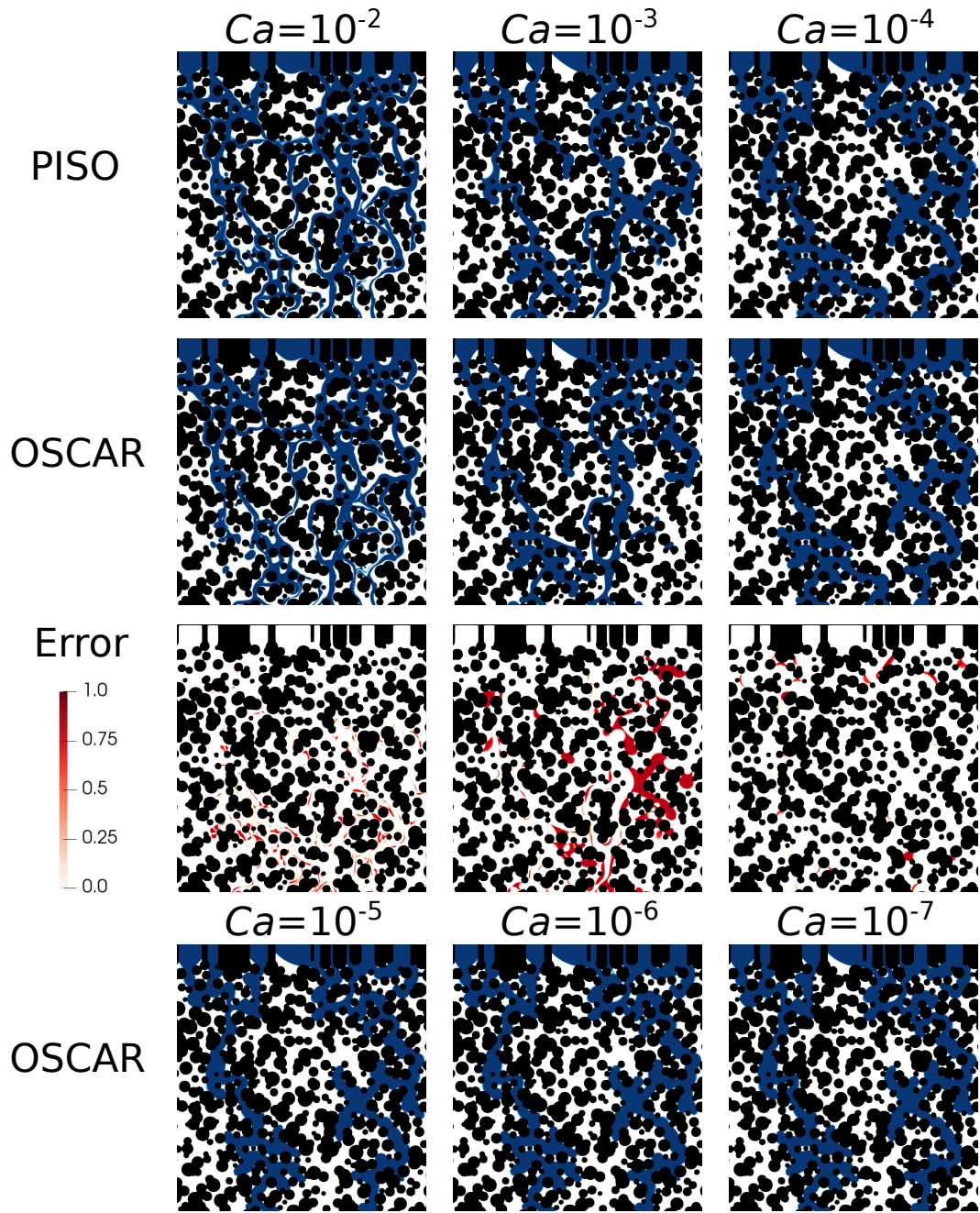}
\caption{Phase distribution at end of simulation during water drainage in an oil-wet micromodel at various capillary numbers. The simulations for $Ca\geq10^{-4}$ are performed with PISO and OSCAR, and the local splitting error in phase volume fraction is shown. The simulations for $Ca<10^{-4}$ cannot be performed with PISO due to high computational cost and only the OSCAR results are shown\label{fig:microSat}}
\end{center}
\end{figure}

Fig. \ref{fig:microSat} shows the phase distribution at the end of each simulation obtained with PISO and OSCAR at $Ca=~10^{-2}$, $10^{-3}$ and $10^{-4}$. PISO and OSCAR shows the same transition between viscous dominated regime ($Ca=10^{-2}$), where the displacement is mostly compact, and capillary dominated ($Ca\leq10^{-4}$), where capillary fingers develop. The local splitting error in phase volume fraction is also shown. At $Ca=10^{-2}$, the error is located in the wetting films, but since the capillary force is dominated by the viscous force, the error does not lead to a significant difference in the invasion pattern. At $Ca=10^{-4}$, the displacement is strongly capillary dominated and the splitting error is low. We observe that the error is larger in the transition between the regimes, i.e. for $Ca=10^{-3}$. The relaxation of the coupling between viscous and capillary forces leads to a small but not insignificant difference in the invasion pattern.

\begin{table}[!b]
\renewcommand\arraystretch{1.5}
\centering
\begin{tabular}{c||c|c||c|c||c}
\hline
 $Ca$ & \multicolumn{2}{c||}{CPU time (hour) } & \multicolumn{2}{c||}{Residual oil saturation}
 & Splitting error \\
    & PISO & OSCAR & PISO & OSCAR &  \\
  \hline
   $10^{-2}$ & \color{blue} 0.56 & 1.12 & 0.433 & 0.434  &  0.067  \\
 $10^{-3}$ & \color{blue} 1.56 & 2.51  & 0.471 & 0.481 & 0.12 \\
 $10^{-4}$ & \color{blue} 16.8 & 18.0 & 0.465 & 0.464  &  0.025  \\
 $10^{-5}$ & 165$^*$ & \color{blue} 63.2& $-$  & 0.456 & $-$ \\
 $10^{-6}$ & 1650$^*$  & \color{blue} 63.9 & $-$  & 0.454  & $-$  \\
 $10^{-7}$ & 16500$^*$  & \color{blue} 64.5  & $-$ & 0.454  & $-$ \\
 \end{tabular}
 \caption{Comparison of CPU time and residual saturation obtained with PISO and OSCAR and evolution of the splitting error (Eq. (\ref{L2split})) for the simulation of water drainage
 in 2D oil-wet micromodel at various capillary numbers.\label{Table:drainage}}
\end{table}

Table \ref{Table:drainage} shows the CPU times and residual oil saturations obtained with PISO and OSCAR, and the evolution of the
splitting error with respect to the capillary number. As noted before, the splitting error is largest (=0.12) for $Ca=10^{-3}$ for which the coupling between viscous and capillary forces matters the most and is low for $Ca=10^{-2}$ and $Ca=10^{-4}$. For $Ca=10^{-3}$, the residual saturation obtained with OSCAR and PISO had a small discrepancy of 2\%. The discrepancy for $Ca=10^{-2}$ and $Ca=10^{-4}$ is 0.2\%. The simulations for $Ca=10^{-2}, 10^{-3}$ and $10^{-4}$ were slightly slower with OSCAR, since for each time-step, viscous and capillary relaxation steps need to be solved. However, the simulations for $Ca=10^{-5}, 10^{-6}$ and $10^{-7}$ could not be performed with PISO due to high CPU time, and the time reported in Table \ref{Table:drainage} 
were obtained by extrapolation. These simulations could be performed using OSCAR and the phase distribution at the end of the simulation is shown on Fig. \ref{fig:microSat}. CPU times and residual oil saturations obtained with OSCAR are given in Table \ref{Table:drainage}. OSCAR gives a speed-up compared to PISO for capillary number $Ca\leq10^{-5}$. OSCAR is 2.5 time faster than PISO for $Ca=10^{-5}$, 25 time faster for $Ca=10^{-6}$ and 250 time faster for $Ca=10^{-7}$. The residual saturation for $Ca\leq10^{-5}$ is lower than for $Ca=10^{-4}$ as the capillary fingers are less ramified.

\begin{figure}[!t]
\begin{center}
\includegraphics[width=0.85\textwidth]{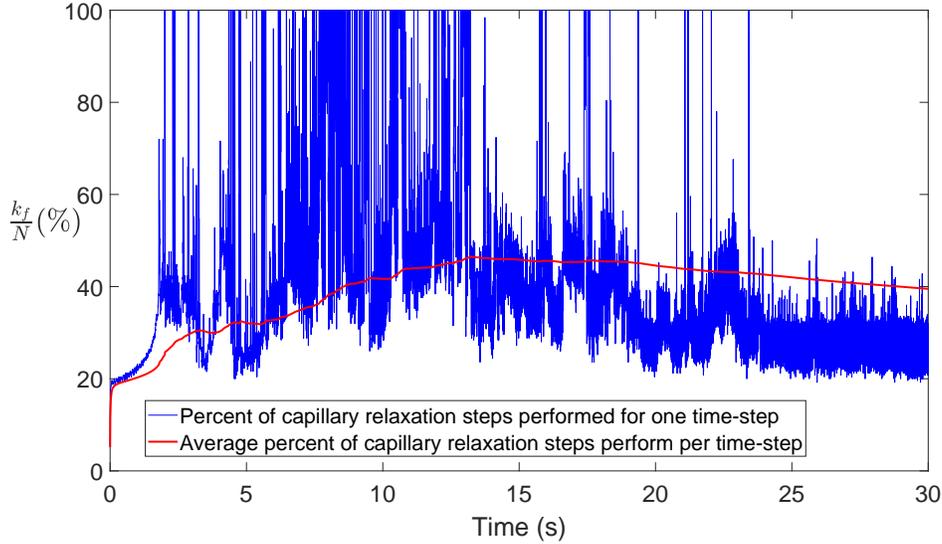}
\caption{Percent of relaxation steps performed for each time-step and average percent of relaxation steps performed as a function of time during water drainage in a 2D oil wet micromodel at $Ca=10^{-5}$.\label{fig:relaxSteps}}
\end{center}
\end{figure}

The percent of relaxation steps performed for each time-step and average percent of relaxation steps performed
for $Ca=10^{-5}$ are plotted in Fig. \ref{fig:relaxSteps}. On average, 39\% of the capillary relaxation steps are performed. Again, the oscillatory nature of the number of capillary relaxation steps in Fig. \ref{fig:relaxSteps} suggests that parasitic currents play a large role in the convergence of our numerical solution and reducing them could further improve the CPU time of OSCAR.

We conclude that OSCAR is capable of simulating accurately drainage in a 2D micromodel for capillary number $Ca\leq10^{-2}$. In particular, it can be used to simulate flow at capillary number $Ca\leq10^{-5}$ in about 60 hours while the CPU time with PISO is more than 165 hours and increases linearly as $Ca$ decreases.

\section{Conclusions}
In this work, OSCAR, a novel time-stepping algorithm for efficient simulation of multiphase flow at low capillary number, is presented. The algorithm uses operator splitting method to separate the viscous drag and the capillary forces. Different time-steps is used for each operator, i.e. based on the CFL number for the viscous drag steps, and based on the minimum of the CFL and of the Brackbill number for the capillary relaxation steps. For one viscous drag step, several capillary relaxation steps are performed until they reach convergence or if they reach the maximum number corresponding to the viscous drag time-step. Convergence, accuracy and efficiency of the method are investigated with three test cases.

In benchmark case 1, the motion of an air bubble in a 2D straight channel filled with dodecane was considered. Convergence of OSCAR with respect to grid size was investigated and an order 1 convergence was observed. The film thickness at the surface of the solid was calculated for $Ca\geq10^{-4}$ and we observed that OSCAR resolves the film with similar accuracy than PISO. Although an additional splitting error of order 0 with respect to the time-step was also observed, this error was smaller than 0.015 for all time-steps and all capillary numbers, and decreased as $Ca$ decreased. For $Ca\leq10^{-4}$, the splitting error was smaller than 0.001 and fast convergence of the capillary relaxation steps led to a large speed-up, so that simulation at $Ca\leq10^{-5}$ could be performed using OSCAR in approximately 20 hours, while these simulations were not performed with PISO due to high computational time.

In benchmark case 2, injection of supercritical CO$_2$ in a 3D constricted channel with aspect ratio equal 3 leading to a snap-off was considered. Simulations were performed at capillary number from $Ca=10^{-4}$ to $Ca=10^{-7}$ with OSCAR and PISO and snap-off was observed in each case. The snap-off time and bubble volume obtained with PISO and OSCAR were similar and the difference decreased with $Ca$. In addition, the splitting error decreased with $Ca$, with an order of approximately 0.7. For $Ca\leq10^{-6}$ the simulation could be performed in approximately 24 hours using OSCAR, while the CPU time with PISO was more than 300 hours and increased linearly as $Ca$ decreases.

In benchmark case 3, water drainage in an 2D oil-wet micromodel was considered. The splitting error was maximal for $Ca=10^{-3}$ and the residual saturation obtained with OSCAR and PISO had a small discrepancy of 2\%, but were almost equal for $Ca\leq10^{-2}$ and $Ca\leq10^{-4}$. For $Ca\leq10^{-5}$, the simulations could be performed with OSCAR in approximately 60 hours, while the CPU time with PISO was more than 160 hours and increased linearly as $Ca$ decreases.

This work paves the way for efficient simulations of multiphase flow at low capillary numbers, which are an essential feature of multiphase reactive transport applications (e.g. CO$_2$ storage, enhanced oil recovery, soil decontamination). Indeed, these applications include processes such as diffusion of ions in water \citep{2021a-Maes} and mineral dissolution \citep{2021-Soulaine} that have time-scales several orders of magnitude larger than the time-scale of capillary waves and at the microscale they can only be modelled using an operator splitting based time-stepping. Our investigation was based on algebraic VOF method, but the principle of splitting the viscous drag and the capillary force can be applied to most multiphase flow methods, i.e. geometric VOF, level-set or phase-field. In particular, there is potential for even more speed-ups since in this work spurious currents may have prevented faster convergence of the capillary relaxation steps. Improving on the convergence criteria using topology-based deficit curvature \citep{2020-Sun} or energy balance \citep{2021-McClure} so that capillary equilibrium can be identified despite parasitic currents could also lead to further speed-up. Splitting methods also isolate the complexity of operators \citep{2017-Iliev} and simplify the application of machine-learning-based acceleration in numerical models \citep{2021-Menke, 2021-Leal}. 

\section*{Acknowledgment}

This work was done as part of the UK EPSRC funded project on Direct Numerical Simulation for Additive Manufacturing in Porous Media (grant reference EP/P031307/1).

\section*{References}
\bibliographystyle{elsarticle-harv}
\bibliography{mybibliography}



\end{document}